\documentclass[fleqn,10pt]{wlscirep}
\usepackage[utf8]{inputenc}
\usepackage[T1]{fontenc}
\usepackage{textcomp}
\usepackage{graphicx}
\usepackage{gensymb}
\usepackage{float}
\title{High-performance magnetostatic wave resonators through deep anisotropic etching of GGG substrates}

\author[1*]{Sudhanshu Tiwari}
\author[1]{Anuj Ashok}
\author[1]{Connor Devitt}
\author[1]{Sunil A. Bhave}
\author[2*]{Renyuan Wang}
\affil[1]{OxideMEMS Lab, Elmore Family School of Electrical and Computer Engineering, Purdue University, West Lafayette, IN 47907 USA}
\affil[2]{FAST Labs, BAE Systems, Inc., Nashua, NH 03060 USA}

\affil[*]{tiwari40@purdue.edu}
\affil[*]{renyuan.wang@baesystems.us}


\begin{abstract}
Microscale resonators are fundamental and necessary building blocks for modern radio communication filters for mobile devices. The resonator's Q factor ($Q$) determines the insertion loss while coupling ($K_t^2$) governs the fractional bandwidth. The product $k_t^2 \times Q$ is widely recognized as the definitive figure of merit for microresonators. Magnetostatic wave resonators based on Yttrium Iron Garnet (YIG) are a promising technology platform for future communication filters. They have shown considerably better performance in terms of $Q$ when compared to the commercially successful acoustic resonators in the $>$7 GHz range. However, the coupling coefficients of these resonators have been limited to $<$3\%, primarily due to the restricted design space imposed by microfabrication challenges related to the patterning of gadolinium gallium garnet (GGG), the substrate material used for growing single crystal YIG. This paper reports novel resonator designs enabled by breakthrough bulk micromachining technology for anisotropic etching of GGG, leading to coupling >8 \% in the 6-20 GHz frequency range. We use the same technology platform to show resonant enhancement of effective coupling, reaching up to 23 \% at 10.5 GHz. The frequency of resonant coupling can be tuned by design during the fabrication process. The resonant coupling results in an unprecedented $k_t^2 \times Q$ figure of merit of 191 at 10.5 GHz and 222 at 14.7 GHz. The technology platform presented in this paper supports both tunable filter architecture and switched filter banks that are currently being used in consumer mobile devices.

\end{abstract}

\begin{document}

\flushbottom
\maketitle

\thispagestyle{empty}

\section*{Introduction}

The backbone of a modern handheld wireless communication system is a Radio Frequency Front-End (RFFE) that selectively transmits and receives signals of interest. Typically, the received RF signal is first filtered using a band pass filter. This filter is one of the most important components of the RFFE because it affects most of the performance metrics of the RFFE, including non-linearity, noise figure, and transceiver power consumption. A typical filter for mobile applications is synthesized using micro-scale resonators. The filter metrics, such as pass-band insertion loss and fractional bandwidth, are limited by the quality factor ($Q$) and coupling constant ($k$) of the resonators utilized in filter synthesis.

The proposed 5G, FR-3 frequency band (7.1-24.2 GHz) is interesting for mobile communication \cite{giribaldi_compact_2024}, Radar systems \cite{ariturk_element-level_2022}, and satellite communications \cite{singh_enhancing_2019}. Notably, there is an emerging requirement for filters within this range to support a bandwidth of $>$ 400 MHz \cite{holma_extreme_nodate}, with even higher demands for satellite communications. Microelectromechanical System (MEMS) resonators based on surface acoustic waves and bulk acoustic waves have dominated the modern RFFE market because of their high performance in the sub-6 GHz frequency range. Current efforts are focused on scaling the acoustic resonators to high frequencies while maintaining the $Q$ and electromechanical coupling. Materials like Lithium Niobate and Scandium doped Aluminum Nitride are being explored for their potential in creating high-coupling acoustic resonators suitable for these advanced applications \cite{giribaldi_compact_2024,esteves2021al0, giribaldi_high-crystallinity_2023, kramer_trilayer_2023, barrera_thin-film_2023,park2021high}.

The acoustic MEMS technology is in the early stage of desired scaling and performance in this frequency range. Filters based on electromagnetic (EM) cavity resonance can provide high performance and can be tuned to the desired frequency\cite{cavity_filter}. However, their size is limited by EM wavelength, which is impractical for mobile applications. Magnetostatic Waves (MSWs) are a subset of spin waves (collective excitations of electron spins) in an ordered magnetic material \cite{stancil2009spin}. The frequency of the MSWs can be tuned by applying an external biasing magnetic field. Yttrium Iron Garnet (YIG) is the preferred material for MSW resonators due to its exceptionally low damping properties. Tunable filters based on the YIG sphere are already mature \cite{roschmann1971yig}. Commercial YIG filters are fabricated by manually aligning coupling transducer loops around a YIG sphere \cite{YIG_Sphere_micro_lambda}. This approach presents significant challenges associated with scaling. Moreover, these filters are bulky and consume high power due to the large tuning current flowing through an electromagnet. Thin film YIG technology offers the advantage of large-scale production and integration with permanent magnets to offer zero quiescent power tuning \cite{du_frequency_nodate}. These attractive properties have fueled renewed interest in YIG thin films for resonators and filters. In this paper, we present a scalable technology based on magnetostatic waves (MSWs) in yttrium iron garnet (YIG) thin films for high $Q$ and high ($k$) resonators suitable to fulfill the needs of future communication systems. 

Single crystal YIG can be grown on a lattice-matched substrate, gadolinium gallium garnet (GGG). This YIG-on-GGG (YoG) platform is being actively utilized to develop tunable resonators \cite{costa_compact_2021, 9108306}, tunable bandstop filters \cite{feng_micromachined_2023}, and tunable bandpass filters \cite{du_frequency_nodate, devitt_edge-coupled_2023}. The YoG platform has also emerged as a promising material platform for other interesting applications, such as magnonics \cite{magnonics_roadmap} and quantum information processing \cite{YIG_quantum_Chumak}. However, microfabrication technologies pertaining to YoG have yet to reach a mature stage. For example, current patterning techniques do not allow complex device structures such as through vias or top-bottom electrodes, which are ubiquitous in Silicon-based materials platforms. Due to these fabrication challenges, most of the YoG devices have the same basic structure where a metallic transducer is patterned on top of a patterned YIG film. This top electrode-only design limits the resonator coupling to < 3 \%; correspondingly, the filter bandwidths are limited to <40 MHz. The resonator coupling is a strong function of the distribution of the RF magnetic field inside the YIG film. The transducers utilized in previous studies utilized metallic transducers on top of the YIG film to generate the excitation RF field. The generated RF field weakly couples to the MSW resonance. Recently, a high coupling ($k_t^2 \sim $ 8 \% @ 18 GHz) resonator was reported through a resonant coupling between a distributed EM resonator and MSW resonance \cite{devitt_distributed_2024} at the cost of size and tunable frequency range.

This paper reports a distinctly different approach to achieving high coupling by employing a transducer that resembles a hairclip-like structure which brings the transducer trace much closer to the ground plane. The device fabrication was made possible because of on new discovery of an anisotropic etching of GGG substrate. This approach results in measured coupling > 8 \% while maintaining a good quality factor in the 6-20 GHz frequency range. We combine the hairclip approach with the resonant approach to achieve a maximum coupling of $\sim$ 23 \% at 10.5 GHz and 17 \% at 14.7 GHz. The coupling enhancement is achieved without any size increase or loss in Q-factor. In order to achieve this device topology, a new microfabrication methodology was developed, which involves thinning and deep etching (up to 100 $\mu$m) of the GGG substrate. This new microfabrication technology enables high-performance resonators for future communication filters. It also opens the pathway of integrating the YoG material platform to devise new designs of other coupled systems for quantum information science. In this regard, coupling magnons with microwave photons\cite{baity2021strong,xu2022strong} and phonons \cite{kounalakis2023engineering,bozhko2020magnon,potts2023dynamical,schlitz2022magnetization,an2020coherent} is an active research area. While this paper focuses on developing resonators for communication filter applications, we show an example of avoided anti-crossing between magnons and microwave photons at room temperature (see supplementary \ref{anticrossing}). With appropriate superconducting metals, compact coupled systems can be developed by utilizing our fabrication technology to work at low temperatures for quantum applications.

\section*{Main}
\subsection*{Hairclip-YIG-on-GGG resonator}

\begin{figure}[H]
\centering
\includegraphics[width=1\linewidth]{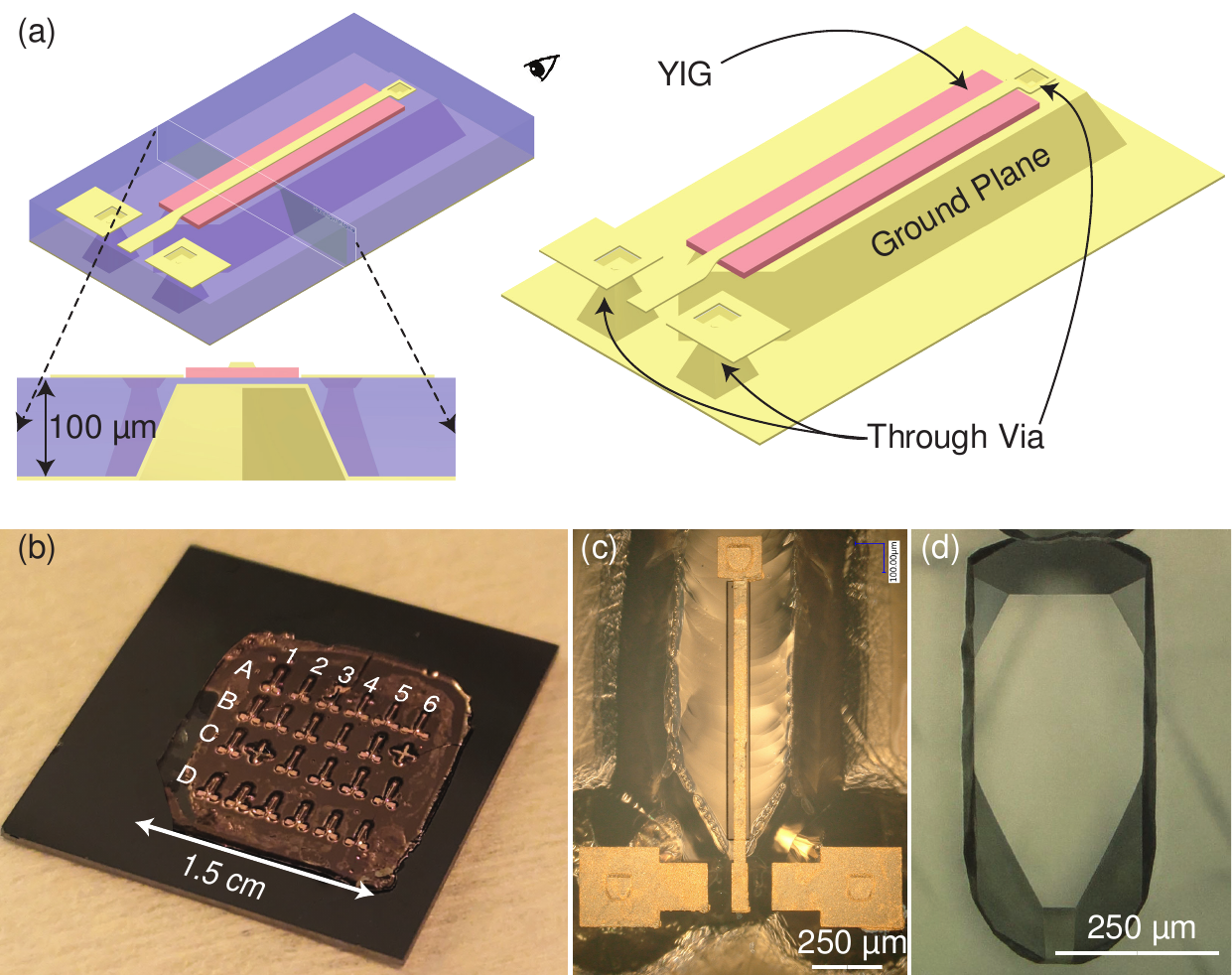}
\caption{\textbf{Design and realization of HYG resonator} (a) Three-dimensional rendering images of the HYG resonator device showing different layers and their topologies. The cross-sectional image (seen from the side of the marked eye) shows a 100 $\mu$m thick GGG substrate on which the resonator was fabricated. The GGG layer is made transparent to show different etching features in the device. A third view, where the GGG layer is hidden, clearly shows the ground plane and through GGG vias (b) Chip micrograph of the HYG resonator showing various resonators on a grid. The HYG chip is mounted on a piece of Si wafer for easy handling. (c) Optical image of a hairclip device after successful fabrication showing different features as explained for the 3D schematic. (d) An optical image of a GGG sample 90 $\mu$m deep etching. The image shows the anisotropic etch profile of the GGG substrate.}
\label{fig:1}
\end{figure}

The resonator design utilizes, now commonly available, the YoG material platform. The design of the hairclip-YIG-on-GGG (HYG) resonators is shown in Fig. \ref{fig:1} (a). The resonator has a hairclip-like inductor loop structure where the narrow top transducer metal meets the wide ground plane at the bottom of the GGG substrate. The continuity of top and bottom metals is facilitated through the via holes in the GGG substrate. The structure of the hairclip is realized by etching three holes in the GGG layer—\textit{two} beneath the ground pads of CPW and \textit{one} adjacent to the YIG. These holes are referred to as Through GGG Vias (TGVs) in the manuscript. The thickness of GGG under the YIG layer is $<$8 $\mu$m. The cross-sectional image in Fig. \ref{fig:1}(a) shows a 100 $\mu$m thick GGG substrate on which the resonator was fabricated. The GGG layer is made transparent to show different etching features in the device. The sectional view also shows the deep-etched GGG layer under the YIG layer. A third view of the resonator with a hidden GGG layer clearly shows the hairclip structure with different TGVs. This design supports the efficient excitation of Magnetostatic Forward Volume Waves (MSFW) when an out-of-plane biasing field \cite{stancil2009spin}. We focus on MSFWs in this manuscript because it is easier to design a magnetic biasing circuit for this configuration. Figure \ref{fig:1}(b) shows an image of the HYG chip containing several HYG resonators. Individual devices on the microfabrication mask were designed to have different opening sizes of TGV etching. Different etch window sizes result in variable thicknesses of GGG under the YIG. Different etching windows also result in the unsuccessful formation of TGVs on some devices. The unsuccessful via formation results in the formation of a capacitor, which results in a resonantly coupled HYG resonator that is described in the upcoming sections of this manuscript.

\subsection*{Performance of tunable Hairclip YoG resonators}

\begin{figure}[H]
\centering
\includegraphics[width=\linewidth]{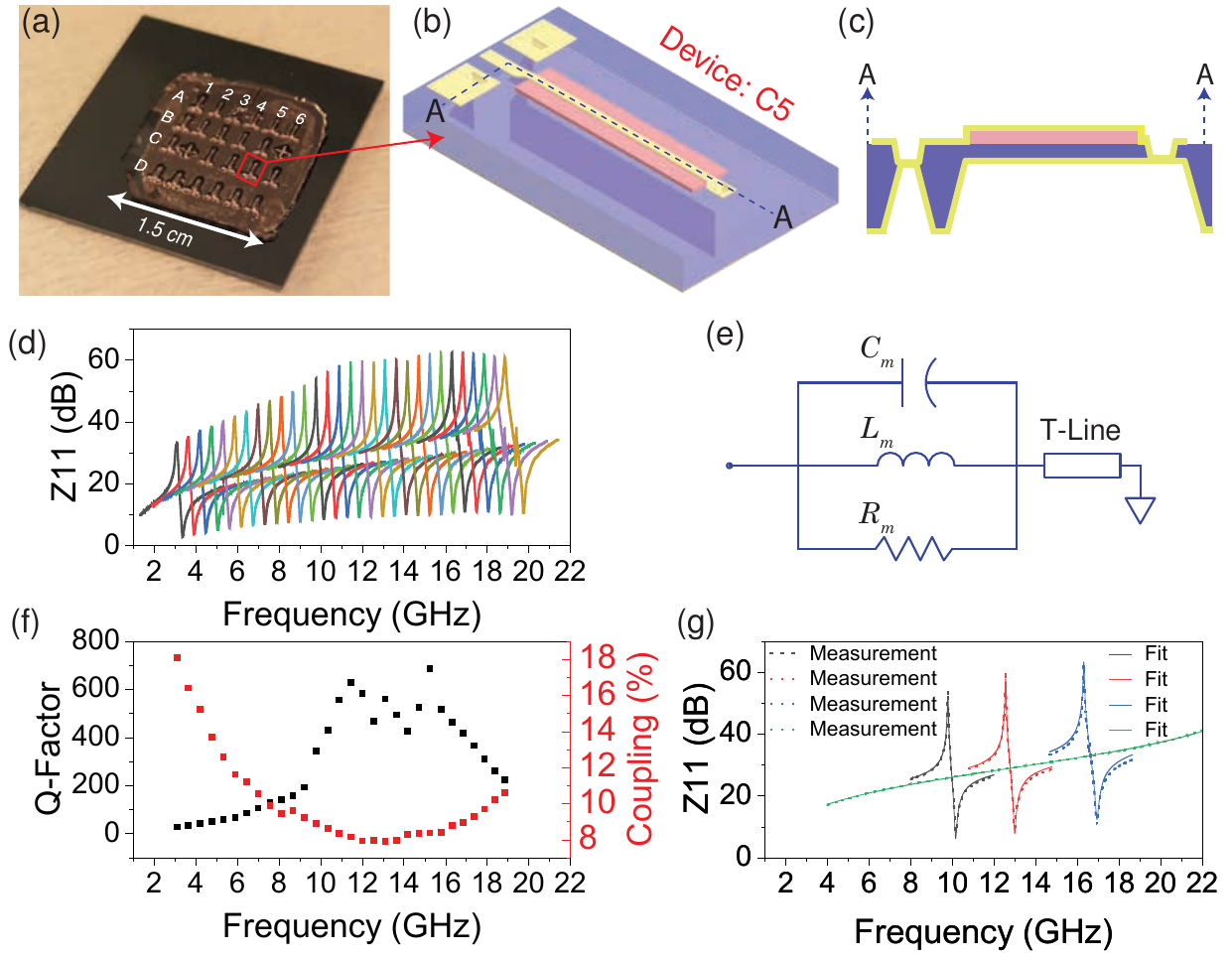}
\caption{\textbf{Measurement results from the Hairclip-YIG-on-GGG (HYG) resonator at room temperature}. (a) A photograph of the fabricated chip showing a grid of resonator devices. (b) A 3D, and (c) a cross-sectional schematic of the HYG resonator showing the formation of the inductive loop around the YIG resonator. (d) The frequency response of the HYG resonator at various bias fields (300 mT to 900 mT). (e) A broadband equivalent circuit model of the HYG resonator showing a Transmission line (representing the transducers) and a parallel RLC branch representing the MSWs (f) Frequency dependence of effective coupling coefficient and quality factor. (f) Circuit fitted response of the HYG resonator at different applied bias fields.}
\label{fig:2}
\end{figure}

An image of the chip with a grid of several resonators is shown in Fig. \ref{fig:2}(a). In this section we discuss the results from a device where we have formed the inductor loop around the YIG resonator through the etched TGVs. A non-magnetic ground-signal-ground (GSG) RF probe is used to measure the reflection parameters of the device. An out-of-plane static biasing field was applied to support MSFV waves in the YIG resonator. Figure \ref{fig:2}(b) shows the measured impedance response of a single HYG resonator at various static bias fields from 300 mT to 900 mT. 

An equivalent circuit model for MSW resonators has been presented in literature \cite{Ethan_ctk_model} and has been utilized by many researchers\cite{9108306, feng_micromachined_2023, du_frequency_nodate}. This circuit model consists of a series resistor ($R_0$) and series indicator ($L_0$) (representing the transducers) with a parallel RLC branch representing MSW waves. This model works perfectly in a narrowband measurement. In this paper, measurement results from a wide frequency range are reported. Hence, the lumped circuit model does not capture the behavior of the device. A broadband circuit model of the resonator utilizing a transmission line is shown in Fig. \ref{fig:2}(e). In this model, the MSW is still modeled as a parallel RLC circuit; however, the transducer is modeled as a transmission line (T-Line). The T-line model is first fitted to the zero-bias data of the resonator. When no bias is applied, the MSW waves do not exist in the resonator, and hence, the fitting parameters of the M-line can be fixed for subsequent MSW fitting. Figure \ref{fig:2} shows measured and fitted impedance response at various bias fields for the combined M-Line MSW response. The parameters of the M-Line are constant for fitting over different bias fields.

The performance metrics of the HYG resonator as a function of resonance frequency are shown in  Figure \ref{fig:2}(d) and (f). From a filter design standpoint, the resonator quality factor ($Q$) and coupling ($k_t^2$) are two main metrics. For a low insertion loss and wide-band filter, high $Q$ and $k_t^2$ are needed. We take the 3dB bandwidth of the highest amplitude MSW resonance for the calculation of the Q-factor. The effective coupling coefficient $k_{t}^2$ was measured using the expression, $ \frac{\pi}{2}  \frac{f_p}{f_s} \times \cot{ \left(\frac{\pi}{2}  \frac{f_p}{f_s}\right)}$, where $f_p$ and $f_s$ are parallel and series resonance frequencies from the measured impedance response. The Q factor of the MSW resonance increases with frequency up to a point (14 GHz) and then starts to decrease. On the other hand, the resonator coupling initially decreases as we go higher in frequency. The initial coupling trend can be understood by intuitive consideration that the equivalent inductance ($L_m$ in Fig. \ref{fig:2}(e)) representing the MSW resonance decreases as we tune it to a higher frequency. This behavior is also expected from reported physics-based equivalent circuit models of spin waves\cite{Ethan_ctk_model}. The successive reduction in $L_m$ results in progressively lower overall energy storage in the MSW resonance as we increase the frequency of MSW waves. The onset of coupling increase is not fully understood. One possible explanation could be the onset of the spurious modes. It is likely that the spurious modes result in higher coupling because of the presence of an additional parallel RLC branch. The initial increasing $Q$ trend has been previously reported\cite{du_frequency_nodate,Q_trend_1987}. To understand the $Q$ drop at higher frequencies, we characterize the transmission line in the absence of any external bias, thereby removing the MSW in the system.  The losses in the Transmission line (1-$|S11|^2$) start to increase rapidly beyond 14 GHz (see supplementary \ref{Sec:Supp-Loss}), which could result in the loss of quality factor. A comparison of performance metrics of the MSW resonators available in the literature is shown in Table \ref{tab:Comparisor_non-resonant}.

\begin{table}[H]
\centering
\caption{Comparison of coupling and FOM from reported MSW resonators in the literature}
\begin{tabular}{|l|c|c|c|r|}
\hline
\textbf{Reference} & \textbf{Frequency}     & $\mathbf{k_t^2}$ (\%)    & $\mathbf{Q}$           & \textbf{Max. FOM} $\left(\mathbf{k_t^2 \times Q}\right)$     \\ \hline
\cite{9108306}       & 3.5 – 7.5 GHz & 0.2       & 1500-5259 & 10.51 @ 4.7 GHz      \\ \hline
\cite{feng_micromachined_2023}    & 6.3 GHz       & 2.8       & 1115      & 31.2 @ 6.3 GHz       \\ \hline
\cite{du_frequency_nodate}      & 3-12 GHz      & 0.2 - 2.4 & 200–1200  & 6 @ 9 GHz            \\ \hline
\cite{costa_compact_2021}  & 4-11 GHz      & -         & 200-350   & 5-10 (4-11 GHz)         \\ \hline
This Work & 3-20 GHz      & 8 - 18    & 100-700  & 58 @ 15.2 GHz        \\ \hline
\end{tabular}
    \label{tab:Comparisor_non-resonant}
\end{table}

\subsection*{Hairclip Resonator: Fabrication}
\begin{figure}[H]
\centering
\includegraphics[width=\linewidth]{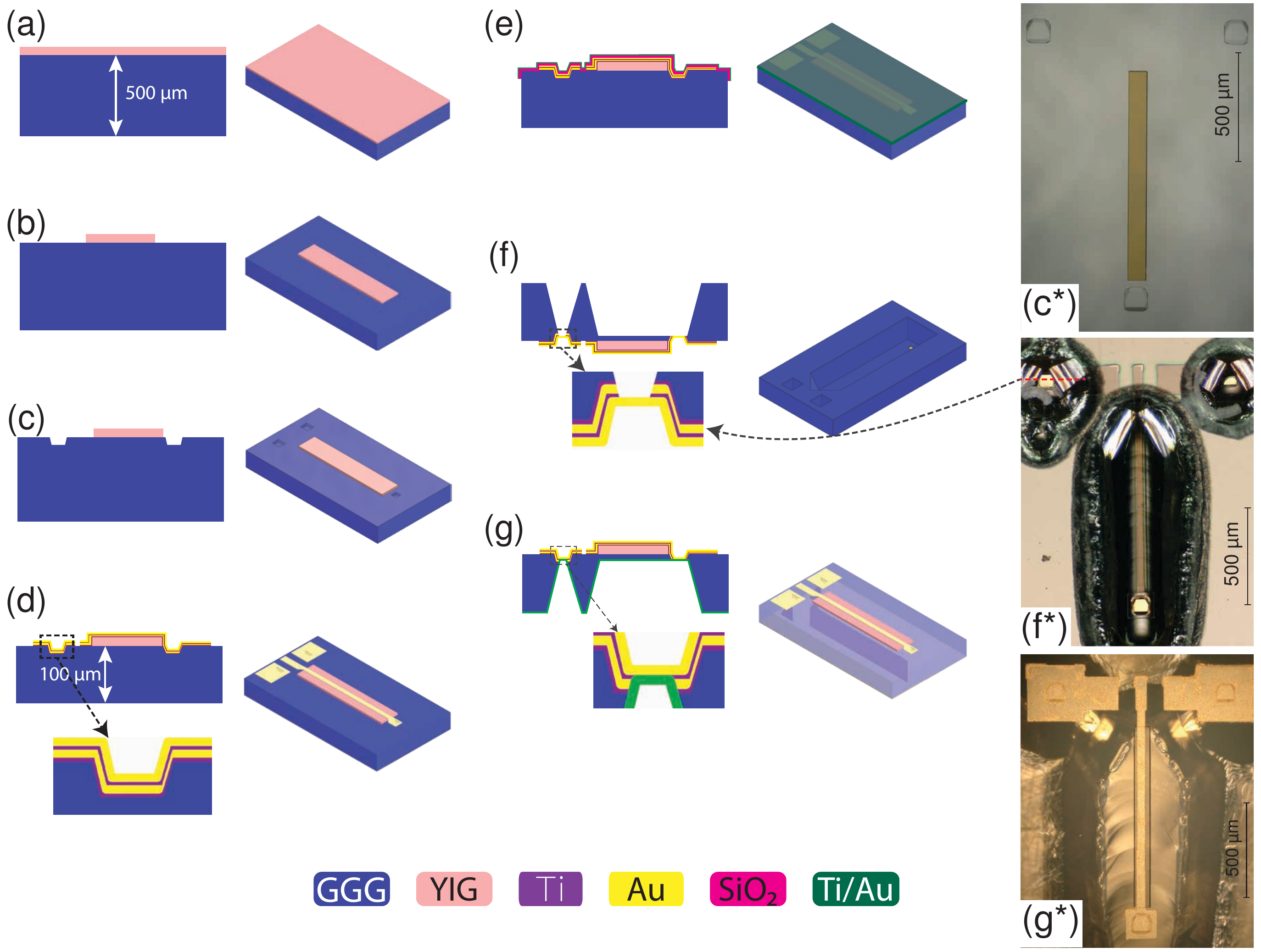}
\caption{\textbf{Fabrication Process Flow for HYG Resonators} (a) Starting YoG material stack. (b) Patterning of YIG layer using ion milling. (c) A shallow (12 $\mu$m) etching of GGG from the top side using H\textsubscript{3}PO\textsubscript{4} @ 160 \degree C. An electroplated Au layer is used as a masking layer for wet etching of GGG. An optical image of the sample after this step is shown in (C*). (d) Patterned electroplating by using a Ti/Au/Ti/Au (10/150/25/150 nm) seed layer that was deposited using the glancing angle deposition method. (e) PECVD deposition of SiO\textsubscript{2} and glancing angle deposition of Ti/Au, which is subsequently used as a seed to electroplate gold as a protection layer for the next processing steps. (f) Backside deep etching of GGG using the same process as (c) to land on the topside electrodes in the shallow etched region. The schematic represents the status of the sample after removing the Au/Ti masking layers. The choice of seed layers in step (d) of the process is important in this step. The first Ti adhesion layer is lost during the GGG etching step, leaving the Au/Ti/Au layer. During the striping of Au/Ti masking layers, the exposed Au and Ti layers are removed in the via region, leaving a visible Au layer as shown in the optical image (f*). (g) Glancing angle deposition of Ti/Au see and subsequent electroplating, which serves as a ground plane.}
\label{fig:process_flow}
\end{figure}

The fabrication of the HYG resonator starts with a 3-inch YoG wafer with a 3 $\mu$m thick YIG layer on a 500 $\mu$m thick GGG substrate. We dice the wafer into 1.5 cm $\times$ 1.5 cm pieces. While we choose to process smaller chips to reduce the cost of the experiment, the process is salable to wafer size. The rest of the processing is performed on 1.5 cm $\times$ 1.5 cm pieces. We start by defining the rectangular YIG patterns on the chip using ion milling and subsequent soaking of the chip in H\textsubscript{3}PO\textsubscript{4} @ 70 \degree C to remove the redeposited material during the ion milling process. After the YIG patterning, we deposit seed metal layers of Titanium (Ti) and Gold (Au) (Ti/Au - 10/300 nm) using the glancing angle deposition (GlAD) method. The seed layer is patterned by lithography and wet patterning of Au and Ti layers. This patterned seed layer is used to deposit 1 $\mu$m Au using electroplating. The patterned electroplated Au serves as the hard mask during the GGG patterning in the next step. A shallow (12 $\mu$m) etching of GGG is performed by utilizing phosphoric acid (H\textsubscript{3}PO\textsubscript{4}) at 160 \degree C. The etch rate of GGG is $\sim$ 0.4 $\mu$m/min in H\textsubscript{3}PO\textsubscript{4} at 160 \degree C (see supplementary \ref{Sec:GGG_Etching}). The hard mask layers (Au/Ti) are stripped from the chip through wet chemical etching. The cross-sectional schematic of the chip after this step is shown in Fig. \ref{fig:process_flow}(c), and the corresponding optical image is shown in \ref{fig:process_flow}(c*). The next step is patterned electroplating to define the top metallic transducers. For this step, a seed layer of stack Ti/Au/Ti/Au-10/150/25/150 nm is deposited using the GlAD method. A photoresist pattern was developed on the chip, and patterned electroplating of 8 $\mu$m gold was carried out. A Ti(100 nm) layer was deposited after the electroplating step, and lift-off was carried out. The partial seed layers (Au/Ti/Au-150/25/150 nm) were removed using wet chemical etching with the Ti mask layer. Finally, the Ti seed and Ti mask were stripped using a buffered oxide etch solution, leaving the patterned top transducers. The details of the process can be found in a previously published report\cite{devitt_distributed_2024}. The chip was then thinned to a thickness of 100 $\mu$m. Figure \ref{fig:process_flow} (d) shows the schematic after completing this step. The top side of the chip was deposited with SiO\textsubscript{2} (0.7 $\mu$m using plasma enhanced chemical vapor deposition) and an electroplated Au layer using the GlAD deposited metals (Ti/Au) as the seed layer. This SiO\textsubscript{2}/Ti/Au layer serves as a protection for subsequent processing steps. Next, a patterned seed layer of Ti/Au-10/300 nm is deposited on the backside of the chip, and electroplating is carried out to get a 1 $\mu$m thick Au mask. The chip is then processed in  H\textsubscript{3}PO\textsubscript{4} \degree C at 160 for $\sim$ 4.5 hours with intermittent metrology steps to track the progress of the etching. At this point, we have completely removed the GGG and the Ti adhesion layer under the shallow etched regions (defined in step (c)) while leaving 3-8 $\mu$m GGG under the YIG layer. Subsequently, the top protection layers (Au/Ti/SiO\textsubscript{2}) are stripped using respective etchants. Figure \ref{fig:process_flow} (f) and (f*) illustrate the cross-sectional profile and optical image of the chip after this step.  We emphasize the choice of seed layers (Ti/Au/Ti/Au) for the top metal traces in step (d), as the bottom Au/Ti layers are lost during this stripping process. We deposit blanket Ti/Au layers using GlAD and use electroplating to achieve a 2 $\mu$m thick gold layer, which serves as the ground plane in our measurement.

\subsection*{Resonantly coupled HYG resonators}
\begin{figure}[H]
\centering
\includegraphics[width=\linewidth]{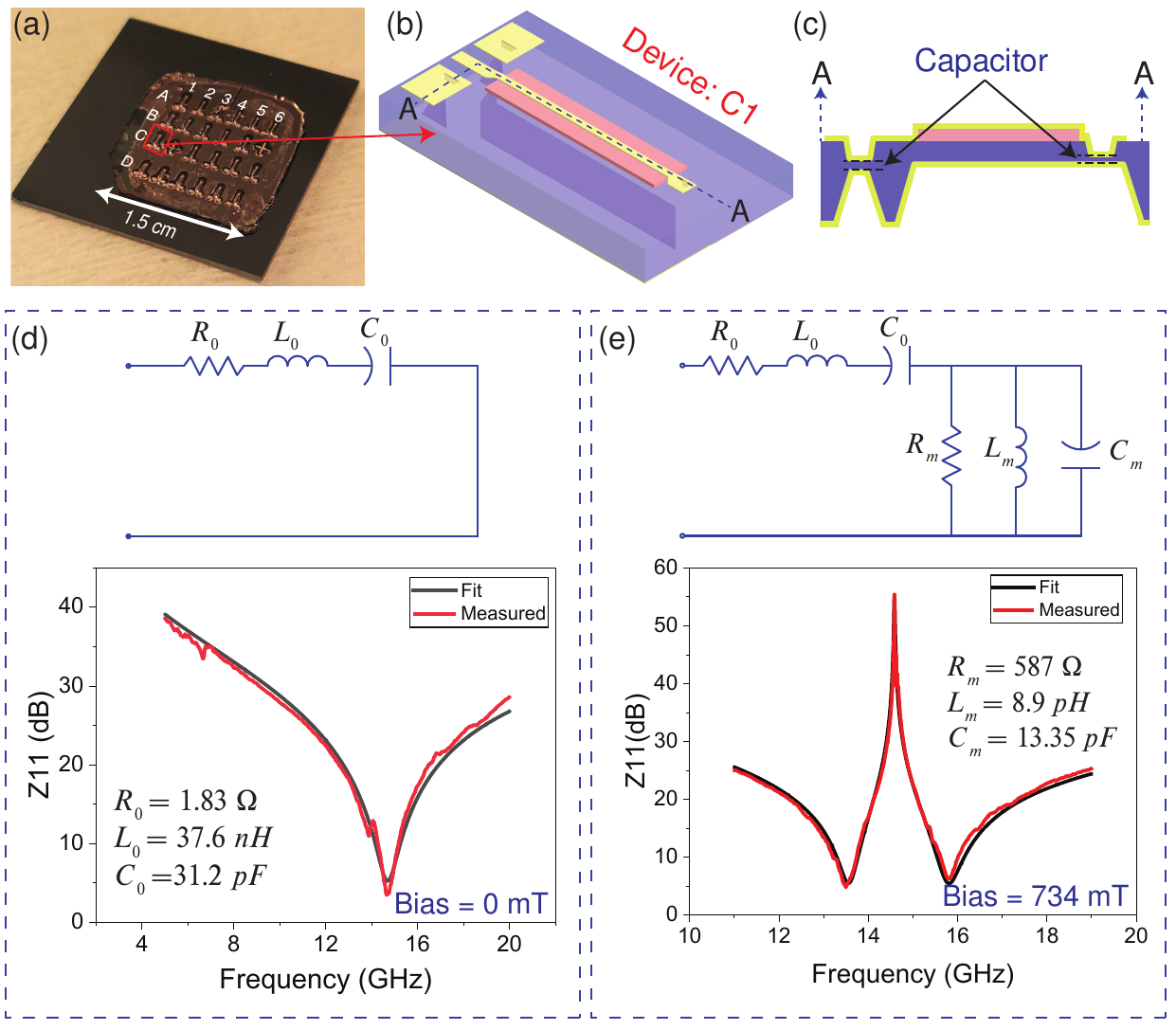}
\caption{\textbf{Measurement results from a resonantly coupled HYG resonator} (a) A photograph of the chip with multiple devices and a marked device, from which the measurement results are plotted in this figure. (b) A 3D schematic of the resonantly coupled hairclip resonator. The 3D schematic is exactly the same as the previously discussed HYG resonator. The difference is only visible in the cross-sectional schematic, as shown in (c). (c) The cross-sectional representation of the resonantly coupled device shows a capacitor formed between the ground place and the topside metal layer. This capacitor results in a self-resonance of the excitation transducer, which is visible in the measured impedance response in the absence of an external bias field. (d) Equivalent circuit and measured impedance response of the device in the absence of any external bias. The measured impedance is fitted to a series LCR circuit, and fitting parameters are fixed for the subsequent fitting of measured MSW resonance. (d) Equivalent circuit and measured response fitted to circuit model. The device response was measured with an external bias of 743 mT.}
\label{fig:C1_main}
\end{figure}

The common techniques of MSW excitation utilize a broadband transducer, such as a Coplaner Waveguide or a microstripline. A resonant coupling between electromagnetic waves and MSW can also be achieved by bringing the EM resonance of the transducer and MSW resonance close to each other. The large wavelength of EM waves in our frequency of interest makes the devices prohibitively large for practical applications. The hairclip fabrication process, on the other hand, presents the opportunity to fabricate a capacitor between the top transducers and the bottom ground plane. This capacitor, coupled with the inductance of the transducer, results in a series LC tank. Since the capacitor is formed in the thickness direction of the device, the LC tank is achieved without incurring any area penalty. Because the etch rate of GGG depends on the size of the etch window, appropriate opening window sizes can be chosen to get variable capacitors. We have variable sizes of etch windows in our layout design for backside GGG etching in step (f) of the fabrication process shown in Fig. \ref{fig:process_flow}. The variable etch rate of GGG results in non-completion of the via etch in some devices. The thickness of the remaining GGG layer determines the capacitance of the capacitor. An LC tank exists from the self-inductance of the transducer and the capacitor formed in the via regions. The cross-sectional schematic of one such device is shown in Fig. \ref{fig:C1_main} (c). When the MSW resonance is close to this LC resonance, a resonant coupling can be achieved. These devices are referred to as resonantly coupled HYG (RHYG) resonators in this manuscript.  

Figure \ref{fig:C1_main} summarizes the working principle of RHYG resonators with measurement results from one device (C1) on the chip. The overall profile of this device is exactly the same as the HYG resonator described earlier. However, the cross-sectional profile (\ref{fig:C1_main}(c)) is different in the sense that the top transducer and bottom ground plane are separated by a thin GGG layer left during the backside etching step. The equivalent circuit and measured impedance response of this device in the absence of an external biasing field are shown in Fig.\ref{fig:C1_main}(d). The measured response fitted to the equivalent circuit and corresponding fitting parameters $R_0$, $L_0$, and $C_0$ are obtained. These parameters are fixed during the subsequent fitting of measured data in the presence of an external bias. The equivalent circuit of the device in the presence of an external biasing field is shown in Fig.\ref{fig:C1_main}(e). A parallel LCR branch is added to the previous circuit to represent the MSW. This circuit is fitted to the measured data with an external bias of 734 mT, and corresponding fitting parameters are reported. The performance results of three resonantly coupled resonator devices are summarized in Table \ref{tab:Resonant_HYG_Summary}. These resonators exhibit unprecedented $k_t^2 \times Q$ figure of merit for any resonator technology in $>$ 10 GHz frequency range.

\subsection*{Applications of HYG resonators}
\begin{figure}[H]
\centering
\includegraphics[width=\linewidth]{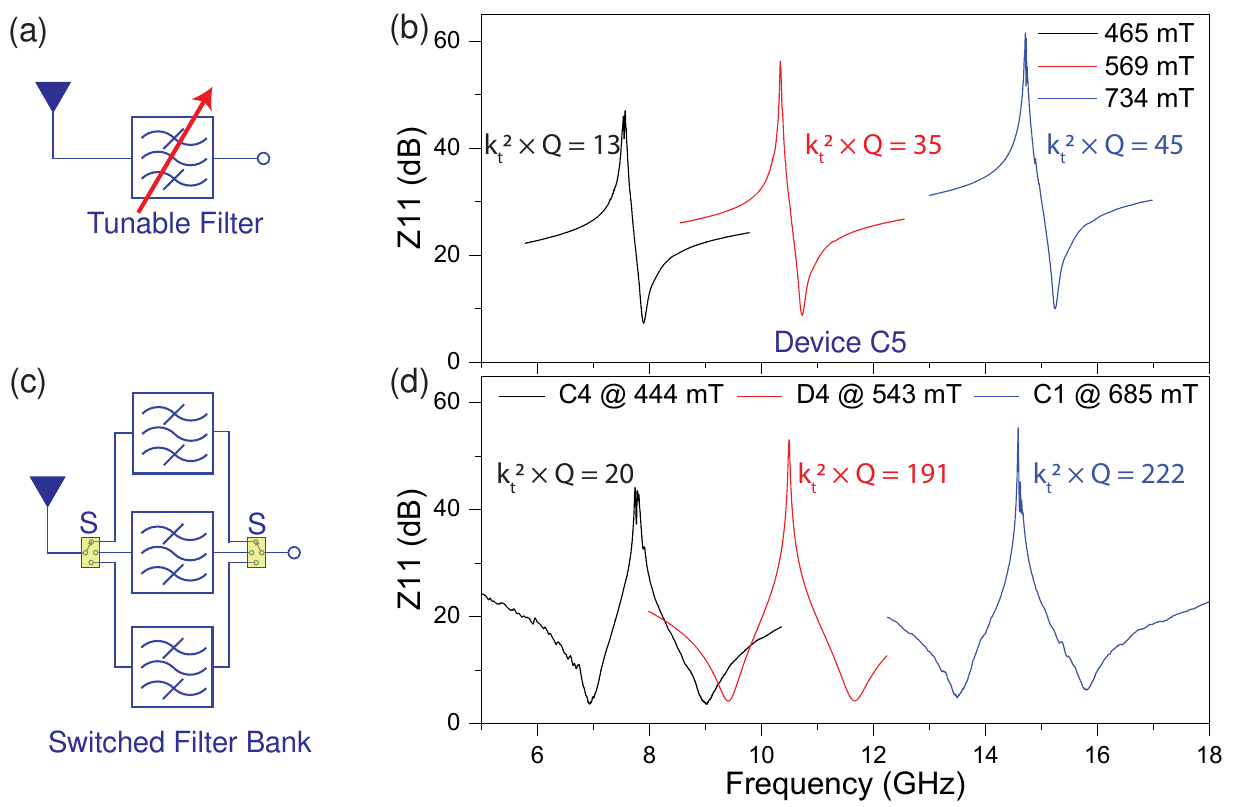}
\caption{\textbf{Bulk GGG etching applied to modern radio technologies} (a) A typical RFFW with a single tunable filter that would employ tunable resonators. (b) Tunable response of a HYG resonator showing octave tunable performance. (c) A typical RFFE with a switched filter bank employing different fixed frequency filters. (d) Measurement results from three different resonantly coupled HYG resonators (C4, D4, and C1) from the fabricated chip. These fixed frequency resonators can be utilized for building switched filter banks.} 
\label{fig:4}
\end{figure}

An RFFE can be assembled with a single tunable filter or a switched filter bank with multiple fixed frequency filters, each covering a distinct frequency band. Switched filter banks are very common in mobile wireless radios, where multiple fixed frequency filters are switched appropriately to receive different frequencies of the signal. The technology platform presented in this paper enables both tunable and switched filters utilizing the same fabrication process. Figure \ref{fig:4}(a) \& (c) illustrate the idea of tunable and switched filters that can be used in a modern front end. Figure \ref{fig:4}(b) shows the tunable response from a resonator device (C5) with three different external bias fields (465 mT, 569 mT, and 734 mT). These are a subset of measured responses that are reported in Fig. \ref{fig:2} (d). Figure \ref{fig:4}(d) shows the impedance responses from three different resonator devices (C4, D4, and C1) that have different series resonances at 7.5 GHz, 10.5 GHz, and 14.7 GHz, respectively. These devices were measured with an external bias to support the MSW near their respective series resonances, which are shown in Fig \ref{fig:4}(d). The tunable hairclip resonators presented in this paper offer significantly better FOM when compared with reported MSW resonators in the literature (see Table \ref{tab:Comparisor_non-resonant}). Furthermore, the RHYG resonators presented in this paper demonstrate performance that is twice as good as the best-reported acoustic resonators\cite{Rouchan_best_FOM} in the $>$10 GHz frequency range.

\begin{table}[H]
\centering
\caption{Summary of measured performance metrics for resonantly couple HYG resonators shown in Fig. \ref{fig:4} (d)}
    \label{tab:Resonant_HYG_Summary}
\begin{tabular}{|c|c|c|c|c|c|}
\hline
Device & External Bias (mT) & Frequency (GHz) & $Q$    & $k_t^2$ (\%) & FOM ($k_t^2 \times Q$) \\ \hline
C4     & 444               & 7.5             & 87   & 23.4     & 20  \\ \hline
D4     & 543               & 10.5            & 839  & 22.8     & 191 \\ \hline
C1     & 685               & 14.7            & 1302 & 17.1     & 222 \\ \hline
\end{tabular}
\end{table}

\section*{Discussion}

Magnetostatic wave resonators offer an alternative technology platform for high-performance communication filters, especially for higher (> 6 GHz) frequency ranges. State-of-the-art thin film YoG resonator technology offers exceptional $f \times Q$ product \cite{9108306}. The two main challenges limiting practical applications of MSW filters are the requirement of bulky and power-hungry biasing magnets and low coupling. Recent reports have shown improvements in scaling biasing magnets' size and power requirements. A compact biasing permanent magnet with temperature compensation has been recently reported for 20 GHz fixed frequency operation\cite{IMS_Temp_Co} with out-of-plane bias. A zero quiescent power biasing assembly has been reported for tunable filter frequency in the 3.4-11 GHz range \cite{du_frequency_nodate}. This paper focused on improving resonator coupling, which compliments the improvement made on the biasing fronts. We achieved the coupling improvement by employing a distinctly different resonator design, which required inventing new microfabrication technology for the patterning of GGG substrates. The anisotropic etching of the GGG substrate that is being reported for the first in this paper paves the way for new designs of magnetostatic wave resonators and filters. The same fabrication methodology allows coupling improvements in two different ways. The hairclip inductive loop around a YIG mesa resonator shows > 8 \% coupling in a wide frequency range (3-20 GHz), which will be suitable for a highly tunable filtering approach. Further enhancement in coupling is achieved through resonant transducers on the same fabrication technology platform. The resonantly coupled resonators have a maximum coupling at a designed frequency, which will be suitable for developing fixed frequency filters with wide fractional bandwidth. The two resonantly coupled resonators presented in this manuscript show a $k_t^2 Q$ FOM of 191 at 10 GHz and 222 at 14.7 GHz, which are the highest reported FOMs > 10 GHz range. The implications of the hairclip resonators lie in their potential to revolutionize communication filters by offering extremely high performance, especially in higher frequency ranges. These resonators address the challenge related to coupling, paving the way for more efficient and tunable filters in wireless communication systems. Moreover, the fabrication technology platform presented in this paper has opened new dimensions for the design of future magnetostatic wave resonators. 

\bibliography{sample}

\begin{thebibliography}{10}
\urlstyle{rm}
\expandafter\ifx\csname url\endcsname\relax
  \def\url#1{\texttt{#1}}\fi
\expandafter\ifx\csname urlprefix\endcsname\relax\def\urlprefix{URL }\fi
\expandafter\ifx\csname doiprefix\endcsname\relax\def\doiprefix{DOI: }\fi
\providecommand{\bibinfo}[2]{#2}
\providecommand{\eprint}[2][]{\url{#2}}

\bibitem{giribaldi_compact_2024}
\bibinfo{author}{Giribaldi, G.}, \bibinfo{author}{Colombo, L.}, \bibinfo{author}{Simeoni, P.} \& \bibinfo{author}{Rinaldi, M.}
\newblock \bibinfo{journal}{\bibinfo{title}{Compact and wideband nanoacoustic pass-band filters for future {5G} and {6G} cellular radios}}.
\newblock {\emph{\JournalTitle{Nature Communications}}} \textbf{\bibinfo{volume}{15}}, \bibinfo{pages}{304}, \doiprefix\url{10.1038/s41467-023-44038-9} (\bibinfo{year}{2024}).

\bibitem{ariturk_element-level_2022}
\bibinfo{author}{Ariturk, G.}, \bibinfo{author}{Almuqati, N.~R.} \& \bibinfo{author}{Sigmarsson, H.~H.}
\newblock \bibinfo{title}{Element-{Level} {Microwave} {Filter} {Integration} in {Fully}-{Digital} {Phased} {Array} {Radar} {Systems}}.
\newblock In \emph{\bibinfo{booktitle}{2022 {IEEE} 22nd {Annual} {Wireless} and {Microwave} {Technology} {Conference} ({WAMICON})}}, \bibinfo{pages}{1--4}, \doiprefix\url{10.1109/WAMICON53991.2022.9786104} (\bibinfo{year}{2022}).

\bibitem{singh_enhancing_2019}
\bibinfo{author}{Singh, V.}, \bibinfo{author}{Ambati, P.~K.}, \bibinfo{author}{Soni, S.} \& \bibinfo{author}{Karthik, K.}
\newblock \bibinfo{journal}{\bibinfo{title}{Enhancing {Satellite} {Communications}: {Temperature}-{Compensated} {Filters} and {Their} {Application} in {Satellite} {Technology}}}.
\newblock {\emph{\JournalTitle{IEEE Microwave Magazine}}} \textbf{\bibinfo{volume}{20}}, \bibinfo{pages}{46--63}, \doiprefix\url{10.1109/MMM.2018.2885674} (\bibinfo{year}{2019}).
\newblock \bibinfo{note}{Conference Name: IEEE Microwave Magazine}.

\bibitem{holma_extreme_nodate}
\bibinfo{author}{Holma, H.}, \bibinfo{author}{Viswanathan, H.} \& \bibinfo{author}{Mogensen, P.}
\newblock \bibinfo{journal}{\bibinfo{title}{Extreme massive mimo for macro cell capacity boost in 5g-advanced and 6g}}.
\newblock {\emph{\JournalTitle{White paper, Nokia Bell Labs}}}  (\bibinfo{year}{2021}).

\bibitem{esteves2021al0}
\bibinfo{author}{Esteves, G.} \emph{et~al.}
\newblock \bibinfo{journal}{\bibinfo{title}{Al0. 68sc0. 32n lamb wave resonators with electromechanical coupling coefficients near 10.28\%}}.
\newblock {\emph{\JournalTitle{Applied Physics Letters}}} \textbf{\bibinfo{volume}{118}} (\bibinfo{year}{2021}).

\bibitem{giribaldi_high-crystallinity_2023}
\bibinfo{author}{Giribaldi, G.}, \bibinfo{author}{Simeoni, P.}, \bibinfo{author}{Colombo, L.} \& \bibinfo{author}{Rinaldi, M.}
\newblock \bibinfo{title}{High-{Crystallinity} 30\% {Scaln} {Enabling} {High} {Figure} of {Merit} {X}-{Band} {Microacoustic} {Resonators} for {Mid}-{Band} {6G}}.
\newblock In \emph{\bibinfo{booktitle}{2023 {IEEE} 36th {International} {Conference} on {Micro} {Electro} {Mechanical} {Systems} ({MEMS})}}, \bibinfo{pages}{169--172}, \doiprefix\url{10.1109/MEMS49605.2023.10052384} (\bibinfo{year}{2023}).
\newblock \bibinfo{note}{ISSN: 2160-1968}.

\bibitem{kramer_trilayer_2023}
\bibinfo{author}{Kramer, J.} \emph{et~al.}
\newblock \bibinfo{title}{Trilayer {Periodically} {Poled} {Piezoelectric} {Film} {Lithium} {Niobate} {Resonator}}.
\newblock In \emph{\bibinfo{booktitle}{2023 {IEEE} {International} {Ultrasonics} {Symposium} ({IUS})}}, \bibinfo{pages}{1--4}, \doiprefix\url{10.1109/IUS51837.2023.10306831} (\bibinfo{year}{2023}).
\newblock \bibinfo{note}{ISSN: 1948-5727}.

\bibitem{barrera_thin-film_2023}
\bibinfo{author}{Barrera, O.} \emph{et~al.}
\newblock \bibinfo{journal}{\bibinfo{title}{Thin-film lithium niobate acoustic filter at 23.5 ghz with 2.38 db il and 18.2\% fbw}}.
\newblock {\emph{\JournalTitle{Journal of Microelectromechanical Systems}}}  (\bibinfo{year}{2023}).

\bibitem{park2021high}
\bibinfo{author}{Park, M.}, \bibinfo{author}{Wang, J.} \& \bibinfo{author}{Ansari, A.}
\newblock \bibinfo{journal}{\bibinfo{title}{High-overtone thin film ferroelectric alscn-on-silicon composite resonators}}.
\newblock {\emph{\JournalTitle{IEEE Electron Device Letters}}} \textbf{\bibinfo{volume}{42}}, \bibinfo{pages}{911--914} (\bibinfo{year}{2021}).

\bibitem{cavity_filter}
\bibinfo{author}{Laplanche, E.} \emph{et~al.}
\newblock \bibinfo{journal}{\bibinfo{title}{Tunable filtering devices in satellite payloads: A review of recent advanced fabrication technologies and designs of tunable cavity filters and multiplexers using mechanical actuation}}.
\newblock {\emph{\JournalTitle{IEEE Microwave Magazine}}} \textbf{\bibinfo{volume}{21}}, \bibinfo{pages}{69--83}, \doiprefix\url{10.1109/MMM.2019.2958706} (\bibinfo{year}{2020}).

\bibitem{stancil2009spin}
\bibinfo{author}{Stancil, D.~D.} \& \bibinfo{author}{Prabhakar, A.}
\newblock \emph{\bibinfo{title}{Spin waves}}, vol.~\bibinfo{volume}{5} (\bibinfo{publisher}{Springer}, \bibinfo{year}{2009}).

\bibitem{roschmann1971yig}
\bibinfo{author}{R{\"o}schmann, P.}
\newblock \bibinfo{journal}{\bibinfo{title}{Yig filters}}.
\newblock {\emph{\JournalTitle{Philips Tech. Rev}}} \textbf{\bibinfo{volume}{32}}, \bibinfo{pages}{322--327} (\bibinfo{year}{1971}).

\bibitem{YIG_Sphere_micro_lambda}
\bibinfo{title}{Technical {Brief} {Details} {YIG}-tuned {Bandpass} and {Band}-reject {Microwave} {Filters}}.
\newblock \bibinfo{note}{\url{https://www.microlambdawireless.com/updates/technology-description-of-yig-tuned-filters/} [Accessed: June 2, 2024]}.

\bibitem{du_frequency_nodate}
\bibinfo{author}{Du, X.} \emph{et~al.}
\newblock \bibinfo{journal}{\bibinfo{title}{Frequency tunable magnetostatic wave filters with zero static power magnetic biasing circuitry}}.
\newblock {\emph{\JournalTitle{Nature Communications}}} \textbf{\bibinfo{volume}{15}}, \bibinfo{pages}{3582} (\bibinfo{year}{2024}).

\bibitem{costa_compact_2021}
\bibinfo{author}{Costa, J.~D.} \emph{et~al.}
\newblock \bibinfo{journal}{\bibinfo{title}{Compact tunable {YIG}-based {RF} resonators}}.
\newblock {\emph{\JournalTitle{Applied Physics Letters}}} \textbf{\bibinfo{volume}{118}}, \bibinfo{pages}{162406}, \doiprefix\url{10.1063/5.0044993} (\bibinfo{year}{2021}).

\bibitem{9108306}
\bibinfo{author}{Dai, S.}, \bibinfo{author}{Bhave, S.~A.} \& \bibinfo{author}{Wang, R.}
\newblock \bibinfo{journal}{\bibinfo{title}{Octave-tunable magnetostatic wave yig resonators on a chip}}.
\newblock {\emph{\JournalTitle{IEEE Transactions on Ultrasonics, Ferroelectrics, and Frequency Control}}} \textbf{\bibinfo{volume}{67}}, \bibinfo{pages}{2454--2460}, \doiprefix\url{10.1109/TUFFC.2020.3000055} (\bibinfo{year}{2020}).

\bibitem{feng_micromachined_2023}
\bibinfo{author}{Feng, Y.}, \bibinfo{author}{Tiwari, S.}, \bibinfo{author}{Bhave, S.~A.} \& \bibinfo{author}{Wang, R.}
\newblock \bibinfo{journal}{\bibinfo{title}{Micromachined {Tunable} {Magnetostatic} {Forward} {Volume} {Wave} {Bandstop} {Filter}}}.
\newblock {\emph{\JournalTitle{IEEE Microwave and Wireless Technology Letters}}} \textbf{\bibinfo{volume}{33}}, \bibinfo{pages}{807--810}, \doiprefix\url{10.1109/LMWT.2023.3267449} (\bibinfo{year}{2023}).
\newblock \bibinfo{note}{Conference Name: IEEE Microwave and Wireless Technology Letters}.

\bibitem{devitt_edge-coupled_2023}
\bibinfo{author}{Devitt, C.}, \bibinfo{author}{Wang, R.}, \bibinfo{author}{Tiwari, S.} \& \bibinfo{author}{Bhave, S.~A.}
\newblock \bibinfo{title}{An {Edge}-{Coupled} {Magnetostatic} {Bandpass} {Filter}} (\bibinfo{year}{2023}).
\newblock \bibinfo{note}{ArXiv:2312.10583 [physics]}.

\bibitem{magnonics_roadmap}
\bibinfo{author}{Chumak, A.~V.} \emph{et~al.}
\newblock \bibinfo{journal}{\bibinfo{title}{Advances in magnetics roadmap on spin-wave computing}}.
\newblock {\emph{\JournalTitle{IEEE Transactions on Magnetics}}} \textbf{\bibinfo{volume}{58}}, \bibinfo{pages}{1--72}, \doiprefix\url{10.1109/TMAG.2022.3149664} (\bibinfo{year}{2022}).

\bibitem{YIG_quantum_Chumak}
\bibinfo{author}{Knauer, S.} \emph{et~al.}
\newblock \bibinfo{journal}{\bibinfo{title}{Propagating spin-wave spectroscopy in a liquid-phase epitaxial nanometer-thick yig film at millikelvin temperatures}}.
\newblock {\emph{\JournalTitle{Journal of Applied Physics}}} \textbf{\bibinfo{volume}{133}} (\bibinfo{year}{2023}).

\bibitem{devitt_distributed_2024}
\bibinfo{author}{Devitt, C.}, \bibinfo{author}{Tiwari, S.}, \bibinfo{author}{Bhave, S.~A.} \& \bibinfo{author}{Wang, R.}
\newblock \bibinfo{title}{A {Distributed} {Magnetostatic} {Resonator}}, \doiprefix\url{10.48550/arXiv.2401.08911} (\bibinfo{year}{2024}).
\newblock \bibinfo{note}{ArXiv:2401.08911 [physics] version: 1}.

\bibitem{baity2021strong}
\bibinfo{author}{Baity, P.~G.} \emph{et~al.}
\newblock \bibinfo{journal}{\bibinfo{title}{Strong magnon--photon coupling with chip-integrated yig in the zero-temperature limit}}.
\newblock {\emph{\JournalTitle{Applied Physics Letters}}} \textbf{\bibinfo{volume}{119}} (\bibinfo{year}{2021}).

\bibitem{xu2022strong}
\bibinfo{author}{Xu, Q.} \emph{et~al.}
\newblock \bibinfo{journal}{\bibinfo{title}{Strong photon-magnon coupling using a lithographically defined organic ferrimagnet}}.
\newblock {\emph{\JournalTitle{arXiv preprint arXiv:2212.04423}}}  (\bibinfo{year}{2022}).

\bibitem{kounalakis2023engineering}
\bibinfo{author}{Kounalakis, M.}, \bibinfo{author}{Kusminskiy, S.~V.} \& \bibinfo{author}{Blanter, Y.~M.}
\newblock \bibinfo{journal}{\bibinfo{title}{Engineering entangled coherent states of magnons and phonons via a transmon qubit}}.
\newblock {\emph{\JournalTitle{Physical Review B}}} \textbf{\bibinfo{volume}{108}}, \bibinfo{pages}{224416} (\bibinfo{year}{2023}).

\bibitem{bozhko2020magnon}
\bibinfo{author}{Bozhko, D.}, \bibinfo{author}{Vasyuchka, V.}, \bibinfo{author}{Chumak, A.} \& \bibinfo{author}{Serga, A.}
\newblock \bibinfo{journal}{\bibinfo{title}{Magnon-phonon interactions in magnon spintronics}}.
\newblock {\emph{\JournalTitle{Low Temperature Physics}}} \textbf{\bibinfo{volume}{46}}, \bibinfo{pages}{383--399} (\bibinfo{year}{2020}).

\bibitem{potts2023dynamical}
\bibinfo{author}{Potts, C.}, \bibinfo{author}{Huang, Y.}, \bibinfo{author}{Bittencourt, V.}, \bibinfo{author}{Kusminskiy, S.~V.} \& \bibinfo{author}{Davis, J.}
\newblock \bibinfo{journal}{\bibinfo{title}{Dynamical backaction evading magnomechanics}}.
\newblock {\emph{\JournalTitle{Physical Review B}}} \textbf{\bibinfo{volume}{107}}, \bibinfo{pages}{L140405} (\bibinfo{year}{2023}).

\bibitem{schlitz2022magnetization}
\bibinfo{author}{Schlitz, R.} \emph{et~al.}
\newblock \bibinfo{journal}{\bibinfo{title}{Magnetization dynamics affected by phonon pumping}}.
\newblock {\emph{\JournalTitle{Physical Review B}}} \textbf{\bibinfo{volume}{106}}, \bibinfo{pages}{014407} (\bibinfo{year}{2022}).

\bibitem{an2020coherent}
\bibinfo{author}{An, K.} \emph{et~al.}
\newblock \bibinfo{journal}{\bibinfo{title}{Coherent long-range transfer of angular momentum between magnon kittel modes by phonons}}.
\newblock {\emph{\JournalTitle{Physical Review B}}} \textbf{\bibinfo{volume}{101}}, \bibinfo{pages}{060407} (\bibinfo{year}{2020}).

\bibitem{Ethan_ctk_model}
\bibinfo{author}{Cui, H.}, \bibinfo{author}{Yao, Z.} \& \bibinfo{author}{Wang, Y.~E.}
\newblock \bibinfo{journal}{\bibinfo{title}{Coupling electromagnetic waves to spin waves: A physics-based nonlinear circuit model for frequency-selective limiters}}.
\newblock {\emph{\JournalTitle{IEEE Transactions on Microwave Theory and Techniques}}} \textbf{\bibinfo{volume}{67}}, \bibinfo{pages}{3221--3229}, \doiprefix\url{10.1109/TMTT.2019.2918517} (\bibinfo{year}{2019}).

\bibitem{Q_trend_1987}
\bibinfo{author}{Murakami, Y.}, \bibinfo{author}{Ohgihara, T.} \& \bibinfo{author}{Okamoto, T.}
\newblock \bibinfo{journal}{\bibinfo{title}{A 0.5-4.0-ghz tunable bandpass filter using yig film grown by lpe}}.
\newblock {\emph{\JournalTitle{IEEE transactions on microwave theory and techniques}}} \textbf{\bibinfo{volume}{35}}, \bibinfo{pages}{1192--1198} (\bibinfo{year}{1987}).

\bibitem{Rouchan_best_FOM}
\bibinfo{author}{Kramer, J.} \emph{et~al.}
\newblock \bibinfo{title}{Thin-film lithium niobate acoustic resonator with high q of 237 and k 2 of 5.1\% at 50.74 ghz}.
\newblock In \emph{\bibinfo{booktitle}{2023 Joint Conference of the European Frequency and Time Forum and IEEE International Frequency Control Symposium (EFTF/IFCS)}}, \bibinfo{pages}{1--4} (\bibinfo{organization}{IEEE}, \bibinfo{year}{2023}).

\bibitem{IMS_Temp_Co}
\bibinfo{author}{Wang, R.} \emph{et~al.}
\newblock \bibinfo{title}{Temperature compensated magnetostatic wave resonator microsystem}.
\newblock In \emph{\bibinfo{booktitle}{2024 IEEE/MTT-S International Microwave Symposium - IMS 2024}} (\bibinfo{year}{2024}).

\bibitem{miller_etch_1973}
\bibinfo{author}{Miller, D.~C.}
\newblock \bibinfo{journal}{\bibinfo{title}{The {Etch} {Rate} of {Gadolinium} {Gallium} {Garnet} in {Concentrated} {Phosphoric} {Acid} of {Varying} {Composition}}}.
\newblock {\emph{\JournalTitle{Journal of The Electrochemical Society}}} \textbf{\bibinfo{volume}{120}}, \bibinfo{pages}{1771}, \doiprefix\url{10.1149/1.2403361} (\bibinfo{year}{1973}).
\newblock \bibinfo{note}{Publisher: IOP Publishing}.

\bibitem{Si_etching_aniso}
\bibinfo{author}{Bean, K.}
\newblock \bibinfo{journal}{\bibinfo{title}{Anisotropic etching of silicon}}.
\newblock {\emph{\JournalTitle{IEEE Transactions on Electron Devices}}} \textbf{\bibinfo{volume}{25}}, \bibinfo{pages}{1185--1193}, \doiprefix\url{10.1109/T-ED.1978.19250} (\bibinfo{year}{1978}).

\bibitem{pal2021high}
\bibinfo{author}{Pal, P.} \emph{et~al.}
\newblock \bibinfo{journal}{\bibinfo{title}{High speed silicon wet anisotropic etching for applications in bulk micromachining: a review}}.
\newblock {\emph{\JournalTitle{Micro and Nano Systems Letters}}} \textbf{\bibinfo{volume}{9}}, \bibinfo{pages}{1--59} (\bibinfo{year}{2021}).

\bibitem{ueda2019review}
\bibinfo{author}{Ueda, J.} \& \bibinfo{author}{Tanabe, S.}
\newblock \bibinfo{journal}{\bibinfo{title}{Review of luminescent properties of ce3+-doped garnet phosphors: New insight into the effect of crystal and electronic structure}}.
\newblock {\emph{\JournalTitle{Optical Materials: X}}} \textbf{\bibinfo{volume}{1}}, \bibinfo{pages}{100018} (\bibinfo{year}{2019}).

\end{thebibliography}

\section*{Acknowledgements}
The device fabrication and measurements were carried out at the Birck Nanotechnology Center, Purdue University. The authors greatly appreciate the insightful discussion with Prof. Pavan Nukala on the crystal structure of GGG. The views, opinions, and/or findings expressed are those of the authors and should not be interpreted as representing the official views or policies of the Department of Defense or the U.S. Government. This work was supported in part by the Air Force Research Laboratory (AFRL) and the Defense Advanced Research Projects Agency (DARPA). The views, opinions and/or findings expressed are those of the authors and should not be interpreted as representing the official views or policies of the Department of Defense or the U.S. Government. This manuscript is approved for public release; distribution A: distribution unlimited.

\section*{Author contributions statement}
R.W. invented and designed the resonator. All authors contributed to defining and developing the fabrication process. S.T. and A.A. developed and optimized unit processes. Final chip fabrication was carried out by S.T., A.A., and C.D. S.T. and C.D. conducted measurements and data analysis under the guidance of R.W. and S.A.B. S.T. and R.W. wrote the manuscript with input from others.

\section*{Supplementary Materials}

\section{Methods}

\subsubsection{Etching of GGG substrate}\label{Sec:GGG_Etching}
\begin{figure}[H]
\centering
\includegraphics[width=\linewidth]{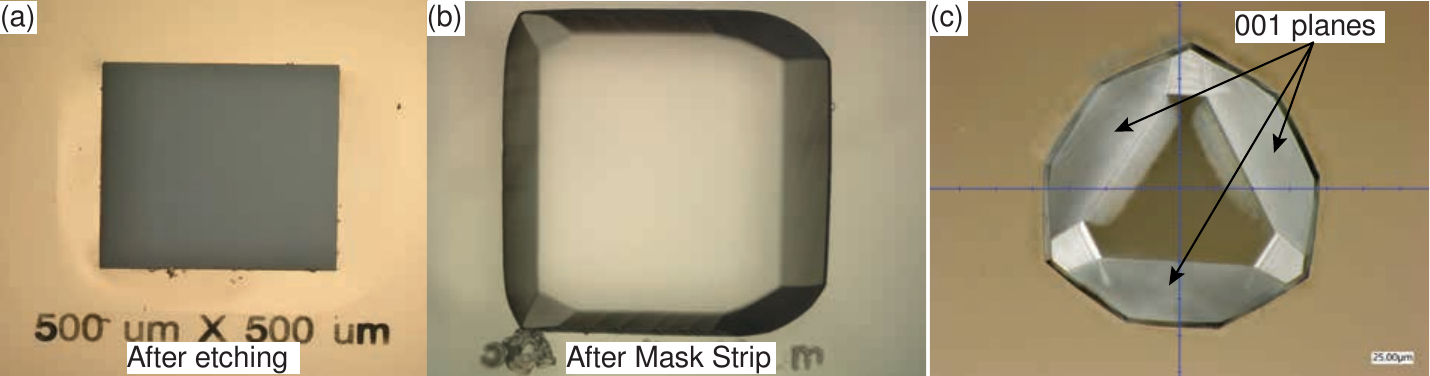}
\caption{\textbf{Deep etching of GGG substrate with gold mask }(a) Optical image of a sample after deep (45 $\mu$m) etching using an electroplated gold mask. The image was captured befor removing the masking gold layer. (c) Optical image of the sample after stripping the gold mask layer. (c) Optical image showing 001 planes of GGG substrate after dep etching of a 20 $\mu$m $\times$ $\mu$m opening area.}
\label{fig:Etch_mask}
\end{figure}

Gadolinium Gallium Garnet (GGG) is the popular substrate for the growth of Yttrium iron Garnet (YIG) because of its lattice match with YIG. The literature is sparse when it comes to etching recipes of GGG in a microfabrication facility. While no reactive ion etching of GGG has been reported in the literature, wet chemical etching has been reported. It has been reported that phosphoric acid (H\textsubscript{3}PO\textsubscript{4}) at elevated temperatures (>140 \degree C) can be used to pattern GGG\cite{miller_etch_1973}. We have used an electroplated gold (Au) layer as a hard mask for phosphoric acid etching of GGG in the 140\degree C -170\degree C range. Figure \ref{fig:Etch_mask}(a) shows an optical image of a 500 $\mu$m $\times$ 500 $\mu$m opening area after 40 $\mu$m deep etching of GGG substrate. Figure \ref{fig:Etch_mask} (b) shows the optical image of the same area after stripping the gold mask layer, showing an anisotropic etch pattern in GGG.

\begin{figure}[H]
\centering
\includegraphics[width=\linewidth]{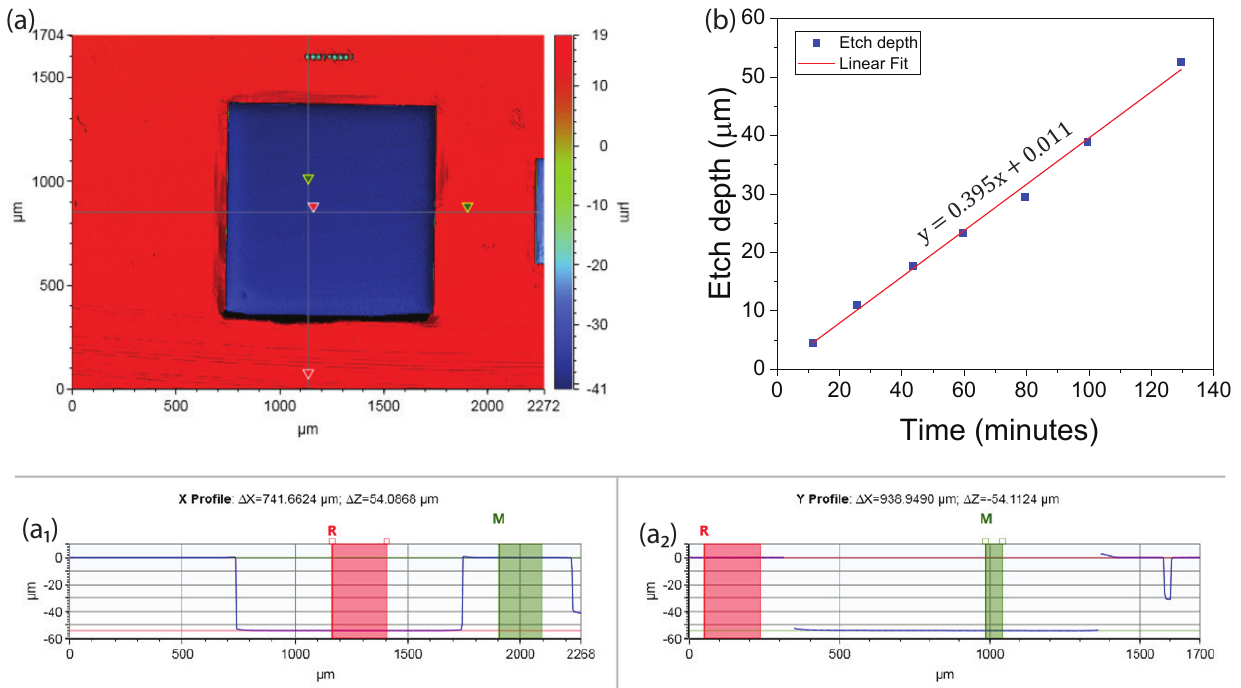}
\caption{\textbf{Deep GGG etching with various masking materials}(a) An optical profilometer measurement of a GGG sample after 54 $\mu$m deep etching. The bottom panels (a\textsubscript{1}) and (a\textsubscript{2}) show depth measurements along horizontal and vertical lines marked in (a). (b) Time vs. etch depth plot for the same sample showing an average etch rate of $\sim$ 0.4 $\mu$m/min for <111> oriented GGG.}
\label{fig:Etch_rate}
\end{figure}

Deep etching of GGG with an electroplated Au mask revealed anisotropic etch rates for different planes of GGG substrate. This observation is similar to the anisotropic etch profiles of Si in KOH or TMAH solution \cite{Si_etching_aniso,pal2021high}. GGG is a member of the garnet family with the general formula A\textsubscript{3}B\textsubscript{2}C\textsubscript{3}O\textsubscript{12} with a cubic crystal structure \cite{ueda2019review}. In GGG crystal (Gd\textsubscript{3}Ga\textsubscript{2}Ga\textsubscript{3}O\textsubscript{12}), Gd atoms occupy dodecahedral sites, while Ga atoms with two different coordination numbers occupy octahedral and tetrahedral sites \cite{ueda2019review}. This complex bond structure results in different etch rates of different crystal planes of GGG. In our experiments, we have used <111> oriented GGG substrates. Figure \ref{fig:Etch_rate} shows an etch depth measurement using an optical profilometer after 54 $\mu$m deep etching into GGG. The etch depth was measured at various time steps during this process to calculate the etch rate. The time vs. etch depth plot is shown in \ref{fig:Etch_rate} (b), showing an average etch rate of $\sim$ 0.4 $\mu$m/min for <111> oriented GGG.

\subsection{Additional Measurement Results}

\subsubsection{Smith plot of HYG resonator}

\begin{figure}[H]
\centering
\includegraphics[width=\linewidth]{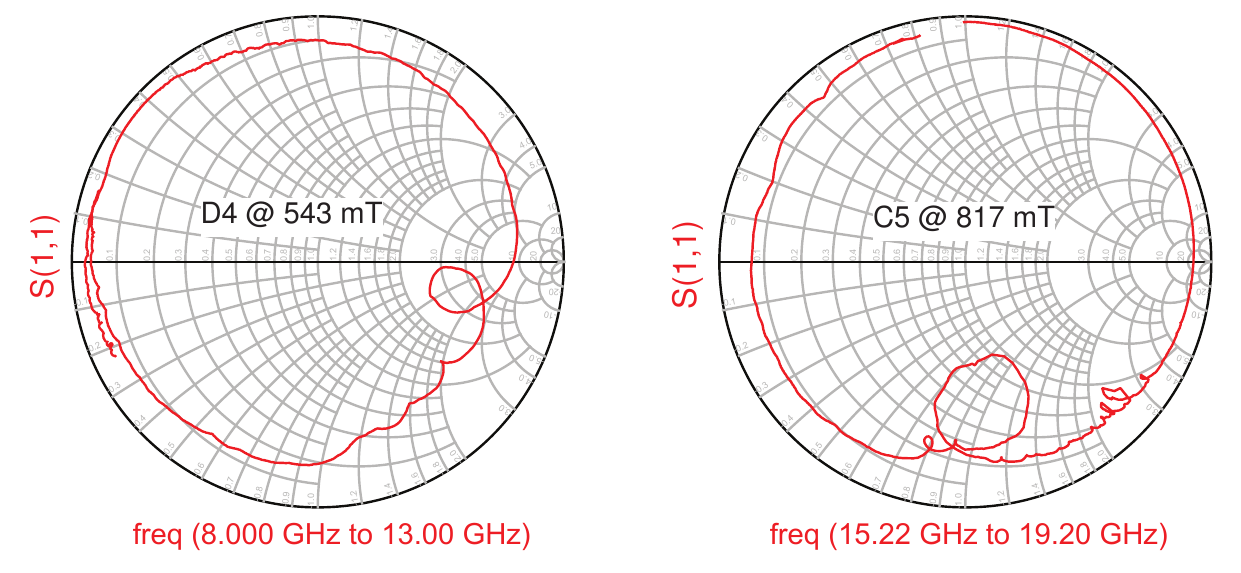}
\caption{\textbf{Resonator performance: Q-Circle}}
\label{fig:Q-circle}
\end{figure}

Resonator reflection on the Smith plot, also termed as "Q-Circle," is shown in Fig. \ref{fig:Q-circle}. Q-Circle plots are useful in understanding the behavior of resonators in terms of spurious modes and deviation from ideal behavior. An ideal resonator would trace the circumference of the Smith chart without any other circle. The additional circles represent the presence of multiple MSW modes. 

\subsubsection{Losses in transmission line}\label{Sec:Supp-Loss}
\begin{figure}[H]
\centering
\includegraphics[width=\linewidth]{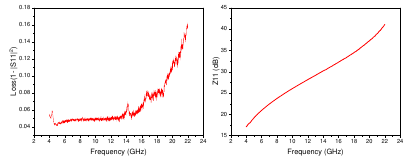}
\caption{\textbf{Reflection measurement results at zero external bias field}(a) Transmission line losses $1-|S11|^2$ in the hairclip resonator structure. (b) Impedance response of the resonator without any external bias.}
\label{fig:Loss}
\end{figure}

We measure the losses in the transmission line of the hairclip resonator by carrying reflection measurements without applying any external magnetic biasing field. The losses in the T-Line calculated as $1-|S11|^2$ are shown in Fig. \ref{fig:Loss}(a). We can see that the losses increase rapidly beyond 14 GHz, resulting in a loss of overall Q-factor in our measurement.

\subsubsection{Additional Resonantly Coupled HYG Resonators}
A measurement result from device C1 on the chip (which is a resonantly coupled HYG resonator) is shown in Figure \ref{fig:C1_main}. The enhancement of coupling is limited near the self-resonance of the transducer. This resonance, however, can be tuned by two methods:- first, by controlling the gap between the top transducer and the ground plane, and second, by controlling the number of capacitors formed during the GGG etch process. The measurement results in Fig. \ref{fig:C1_main} are from a device where all three via regions on the device were capacitively coupled. In other words, none of the via holes were successfully etched during the etch process.

\begin{figure}[H]
\centering
\includegraphics[width=\linewidth]{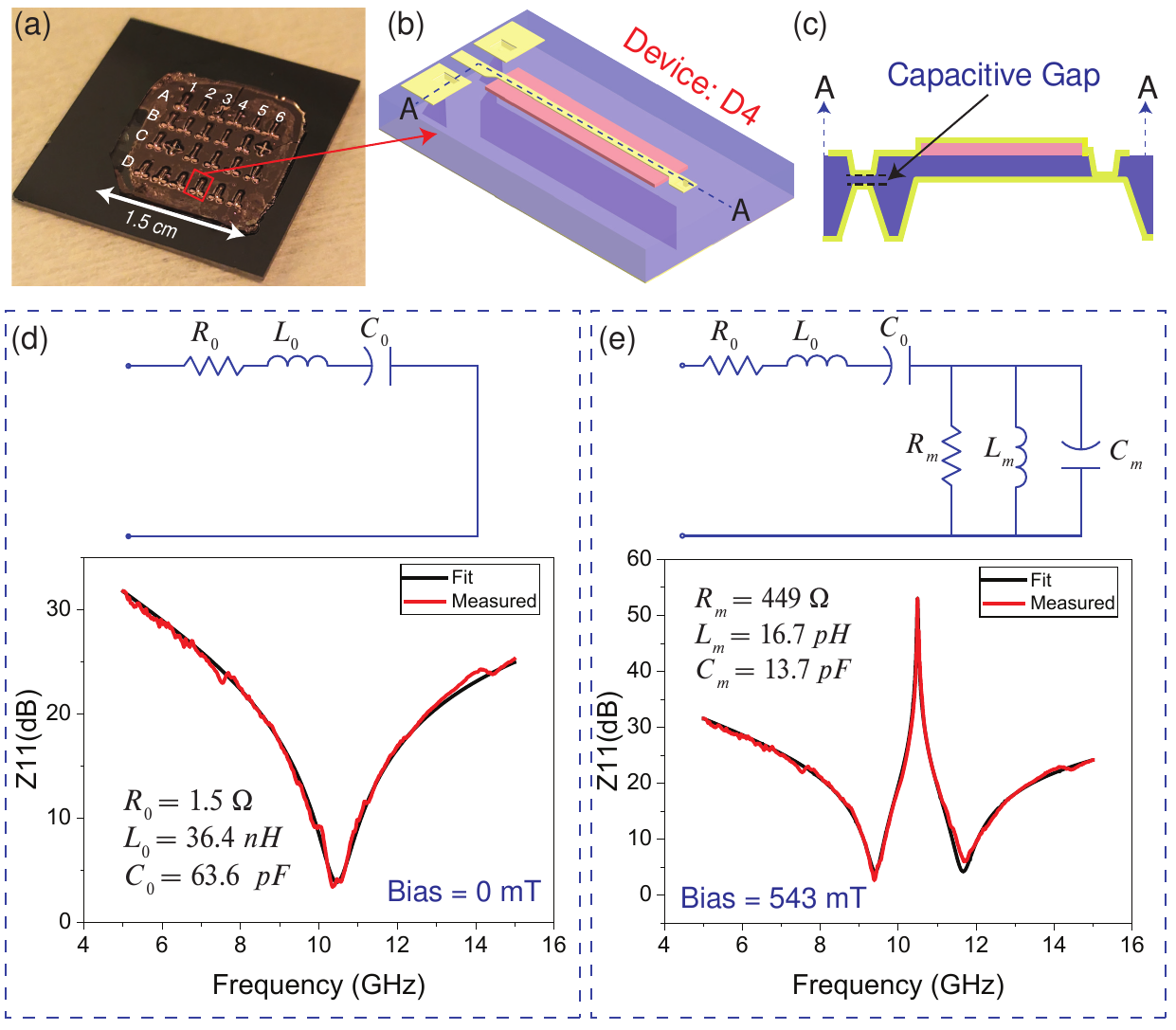}
\caption{\textbf{Measurement results from resonantly coupled HYG.} (a) An image of the chip showing the grid of devices and marking the device from which the results are shown in this figure. (b) 3D schematic of the device. (c) A cross-sectional schematic of the device. (d) The measured impedance response of the device without any external bias and corresponding series LCR circuit fitting parameters. (e) The measured response of the device with an external bias of 543 mT and corresponding fitting parameters.}
\label{fig:D4_main}
\end{figure}

In this section, we show measurement results from another (D4 on the grid shown in Fig. \ref{fig:D4_main}(a)) resonantly coupled HYG resonator. The self-resonance of the transducer for this device is located near 10 GHz (see Fig. \ref{fig:D4_main}(d)). The equivalent circuit and impedance response of one such device on the chip is shown in Fig. \ref{fig:D4_main}(d) without an external bias. As expected, a series LCR circuit fits the response of the resonator with external bias. When an external bias is applied, a parallel LCR circuit representing the magnetostatic wave appears. The measured impedance response fits the equivalent circuit, as shown in Fig. \ref{fig:D4_main}(e). 

\subsubsection{Definition of coupling coefficient for resonantly coupled HYG resonators}\label{Coupling_definition}
 The frequency response resonantly coupled HYG resonator exhibits distinctly different forms when compared to the HYG resonators discussed previously. This leads to a different definition of coupling. In a broadband sweep, the frequency response will always have a high impedance peak ($f_m$) corresponding to MSW resonance and two low-impedance anti-resonances $f_{s1}$ and $f_{s2}$. In this scenario, we use the coupling coefficient as follows, 

\begin{equation}
    k_t^2 = \frac{\pi}{2}  f_{\text{ratio}} \times \cot{ \left(\frac{\pi}{2}   f_{\text{ratio}}\right)}
    \label{eq:Coupling}
\end{equation}

\noindent
where the ratio $ f_{\text{ratio}}$ is either $\left( \frac{f_m}{f_{s2}} \right)$ or $\left( \frac{f_{s1}}{f_{m}} \right)$ (). The ratio is calculated from the minimum separation between resonance and anti-resonance frequencies, which is a number <1.

\subsubsection{Photon Magnon Anti-Crossing}\label{anticrossing}
\begin{figure}[H]
\centering
\includegraphics[width=0.49\linewidth]{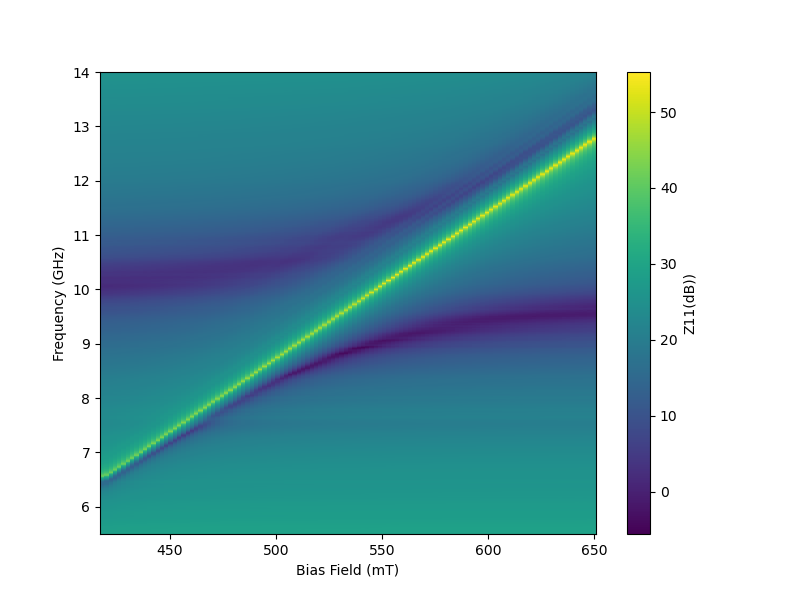}
\includegraphics[width=0.49\linewidth]{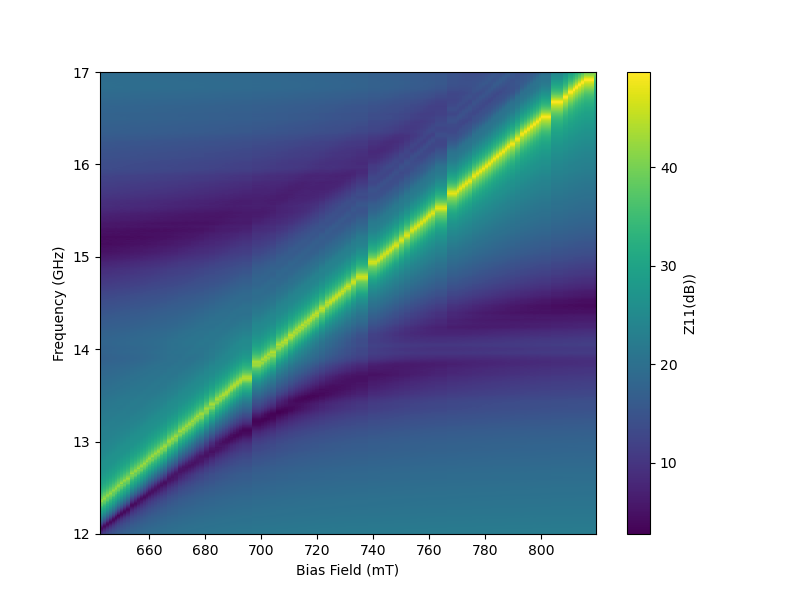}
\caption{\textbf{Photon magnon anti-crossing results} (a) A heatmap plot of impedance response from the RHYG resonator with self-resonance near 10 GHz. (b) A heatmap plot of impedance response from the RHYG resonator with self-resonance near 15 GHz. Both responses show a characteristic avoided photon-magnon anti-crossing at their respective self-resonances.}
\label{fig:6}
\end{figure}

As described earlier, the ability to design an electrical resonance on the device allows for the opportunity to study the coupling of photons and magnons. While this paper is not focused on this study, we show the characteristic coupling response from our devices. Figure \ref{fig:6} shows the heatmap of measured impedance over frequency and applied bias field. The characteristic anti-crossing between electrical and magnon resonance can be clearly seen in the plot.

\subsubsection{Measurement Setup}\label{Measurement_setup}

\begin{figure}[H]
\centering
\includegraphics[width=1\linewidth]{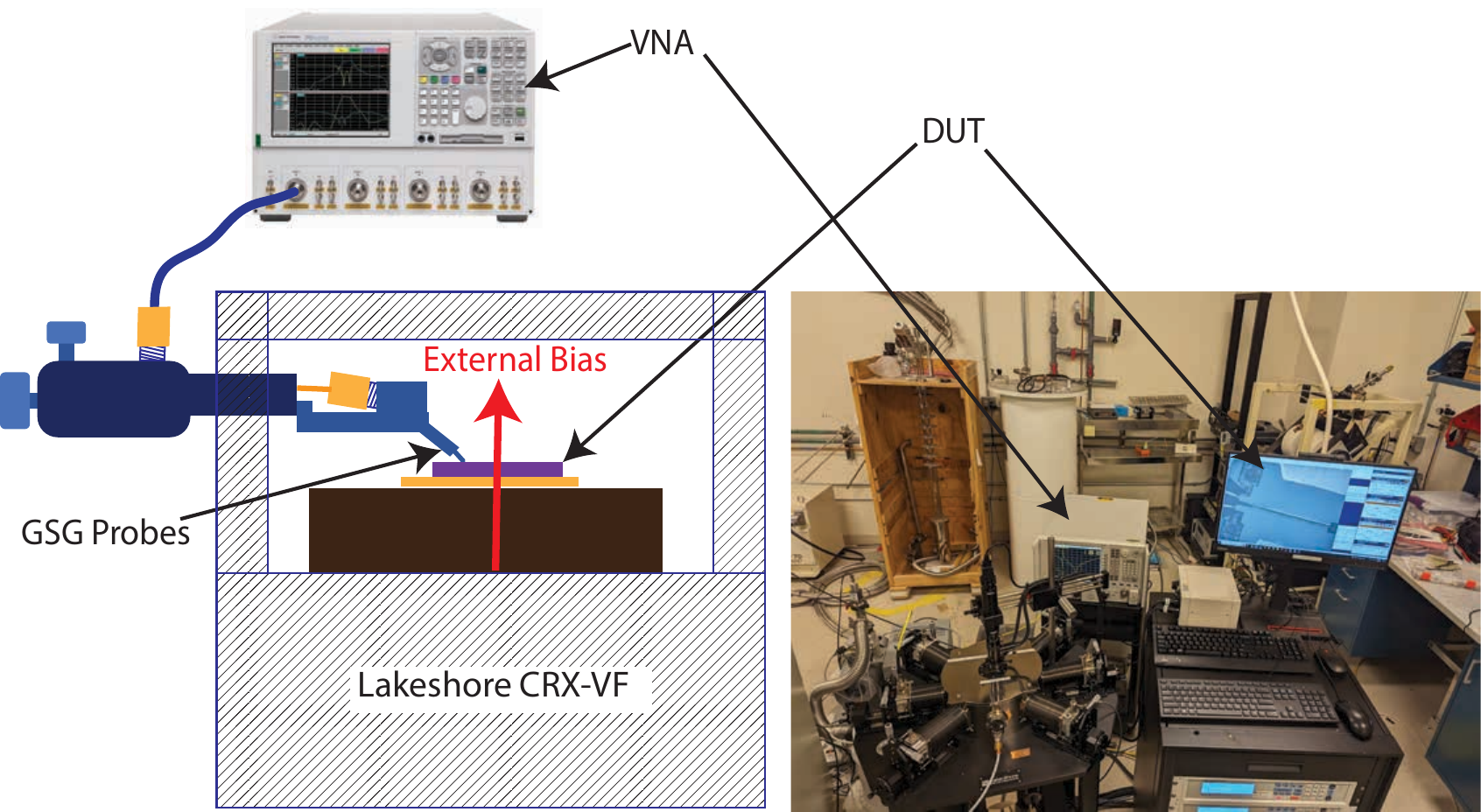}
\caption{\textbf{Measurement Setup:} A lakeshore CRX-VF probe station with a vertical magnetic field and RF measurement capabilities. }
\label{fig:measurement_setup}
\end{figure}

An RF probe station with out-of-plane magnetic biasing capability (LakeShore \textregistered CRX-VF) in combination with a vector Network Analyzer (VNA) from Agilent N5230A was used to measure the RF characteristics of the resonators. Figure \ref{fig:measurement_setup} shows a schematic representation of the measurement setup. All measurements were carried out after calibrating the VNA up to the GSG probe tips following OSL (Open-Short-Load (50 $\Omega$)) calibration standards. The reflection parameters were measured by applying an input power of -15 dBm while a static bias was applied. The reflection parameter (S11) was measured with 20001 points with an average of \textit{three} sweeps. The measured S11 was converted to Z11 using the 50 $\Omega$ input impedance standard. The paper reports calculated quality factors and effective coupling coefficients. The calculation methodology is described in the manuscript. 

\end{document}